\documentclass[proceedings, preprint]{rmaa}

\usepackage{paralist}

\usepackage{psfrag,color}

\SetYear{2007} \SetConfTitle{The Nuclear Region, Host Galaxy, and
Environment of \\Active Galaxies}

\title{Accretion Disks and the Nature and Origin of AGN Continuum Variability}

\author{C. Martin Gaskell\altaffilmark{1}}

\altaffiltext{1}{Astronomy Department, University of Texas, Austin,
TX 78712 (gaskell@astro.as.utexas.edu).}

\shortauthor{Gaskell}
\shorttitle{AGN Variability}

\listofauthors{C.~M. Gaskell}
\indexauthor{Gaskell, C.~M.}

\abstract{Theory and observations of the dominant thermal continuum
emission in AGNs are examined.  After correction for reddening, the
steady state AGN optical--UV spectral energy distributions (SEDs)
are very similar. The SEDs are dominated energetically by the ``big
blue bump'' (BBB), but this bump never shows the $\nu^{+1/3}$
spectrum predicted for a standard thin accretion disk with a
$r^{-0.75}$ radial temperature gradient.  Instead, the observed
optical--UV SED implies a temperature gradient of $r^{-0.57}$
independent of the thickness of the disk. This means that there is
some flow of heat outwards in the disk.  The disk is large and the
region emitting the optical continuum is as large as the inner
broad-line region (BLR). Because optical variability is seen in all
AGNs on the light-crossing time of the BLR, variations must
propagate at close to the speed of light, rather than on dynamical
timescales. This argues that the energy-generation mechanism is
electromagnetic rather that hydrodynamic. Since the velocities are
near the speed of light, there can be significant local anisotropy
in the emission. The large rapid variations of the BBB imply that
the magnetohydrodynamic energy generation is fundamentally unstable.
Because of the inevitable radial temperature gradient in the
accreting material, {\it different spectral regions come
predominantly from different radii}, and variations in different
spectral regions correspond to variability at different radii. This
explains the frequently observed independence of X-ray and optical
variations, cases of variability at lower energies leading
variability at higher energies, and rapid changes in emission-line
reverberation lags. Some observational tests of the local
variability hypothesis are proposed.}

\resumen{
}

\addkeyword{accretion, accretion disks }

\addkeyword{black hole physics}

\addkeyword{galaxies: active}

\addkeyword{quasars: general}

\addkeyword{galaxies: Seyfert}

\begin{document}
\maketitle

\section{Introduction}

A symposium in honor of Deborah Dultzin's 50th birthday\footnote{The
``60th birthday'' in the symposium title is clearly a mistake.
Deborah can't really have turned 60 because that would have
implications for my own age that I am not sure I am quite ready to
face yet!} is a good time to look at AGN variability because it is a
major area of research in which Deborah has worked (see, for
example, Dultzin-Hacyan et al.\@ 1992, Crenshaw et al.\@ 1996,
Edelson et al.\@ 1996, de Diego et al.\@ 1998, and Ram{\'{\i}}rez et
al.\@ 2003). There is much discussion elsewhere in these proceedings
of variability of the {\it blazar} component of AGN emission (an
area of research in which Deborah and her collaborators have been
particularly heavily involved), but in this paper I am just going to
focus on what I think are some of the fundamental implications of
{\it non}-blazar AGN continuum and continuum variability
observations when confronted with simple accretion theory. For a
more comprehensive overview of the state of our knowledge of AGN
variability across the electromagnetic spectrum the reader see
papers in Gaskell et al.\@ (2006).

\section{The Steady-State Spectrum of the Non-Blazar Emission}

Unfortunately, one cannot see what the SED of an AGN is like simply
by adding up flux-calibrated spectra from one's favorite satellites
because solid-state and atomic absorption features caused by dust
and gas along the line of sight modify the SED. The average
properties of dust in AGNs are modified by the AGN environment so
the extinction curve of AGNs differs somewhat from that of dust in
the solar neighborhood (see Gaskell et al.\@ 2004, Czerny et al.\@
2004, and Gaskell \& Benker 2007).  The {\it observed} SEDs of AGN
show considerable variety, especially in the optical to UV, but we
have argued (see Gaskell et al.\@ 2004) that this is predominantly
due to differing amounts of extinction. Once AGN SEDs are corrected
for extinction, the SEDs in the optical to UV are very similar.  A
mean AGN extinction curve can be found in Gaskell \& Benker (2007).

Sometimes history and conventions get in the way of understanding
what is really going on in a subject.  I believe the way AGN
continua have commonly been graphed is an example of this. It was
recognized over half a century ago that the spectra of strong radio
sources were non-thermal and could be approximated by a $F_\nu
\propto \nu^{\alpha}$ power law in frequency, $\nu$, where $F_\nu$
is the energy per unit frequency, and $\alpha$ is a power-law index.
$F_\nu \propto \nu^{-1}$ is also not a bad approximation of the
overall spectrum from the radio region to the optical in many
radio-loud AGNs.  Since $\alpha$ is, of course, the slope of a line
in a log--log plot, it was only natural to make $\log F_\nu$ vs.\@
$\log \nu$ plots of the the overall SEDs of AGNs.  An example of
this is Fig.\@ 2 of Shields (1978), the first paper to point out
that there are clear signs of thermal emission in AGNs. The
unmistakable impression one gets from such plots (see Fig.\@ 1) is
that the dominant thing is the power law and that there is a modest
bump (the ``big blue bump'', BBB) superimposed on top of it.

With increased interest in the significance of various spectral
components it has now become standard practice to make $\log \nu
F_\nu$ vs.\@ $\log \nu$ plots since these emphasize deviations from
a $F_\nu \propto \nu^{-1}$ power law more clearly. What was formerly
a downwards sloping line in the earlier plots now becomes deviations
from a horizontal line.  This makes various wiggles in the overall
SED clearer, but the impression is still that the dominant thing is
the $F_\nu \propto \nu^{-1}$ power law which has now been
transformed into a horizontal line. A much more informative way of
plotting things is to make a log--{\it linear} plot of $\nu F_{\nu}$
(rather than $\log \nu F_{\nu}$ vs.\@ $\log \nu$).  As Carleton et
al.\@ (1987) point out, such a plot ``is particularly useful in that
an area under the plotted spectrum between two values of $\log \nu$
represents the power actually radiated in that frequency band.''

Figs.\@ 1 and 2 respectively show $\log F_\nu$ vs.\@ $\log \nu$ and
$\nu F_{\nu}$ vs.\@ $\log \nu$ plots  of our best estimate of the
SED of the well-observed AGN NGC 5548 at a single epoch.  The SED
has had the host galaxy emission removed and has been corrected for
reddening both by dust in both the host galaxy and the Milky Way .
Details can be found in Gaskell, Klimek, \& Nazarova (2007). Note
the strikingly different impressions Figs.\@ 1 and 2 create!  Since
areas under the curve in the $\log \nu F_{\nu}$ vs.\@ $\log \nu$
plot represent the power per decade, it can be seen clearly from
Fig.\@ 2 that, rather than being dominated by a $F_\nu \propto
\nu^{-1}$ power law, the spectrum is instead dominated by a broad
peak in the far UV.  The remaining emission is primarily the result
of reprocessing the radiation of the broad far UV peak.  The smaller
bump at $\log \nu \sim 13$ is caused by thermal emission from hot
dust in the torus, and the X-ray emission at frequencies higher than
$10^{17}$ Hz is believed to be due to Comptonization of lower-energy
radiation.

\begin{figure}[!t]
  \includegraphics[width=\columnwidth]{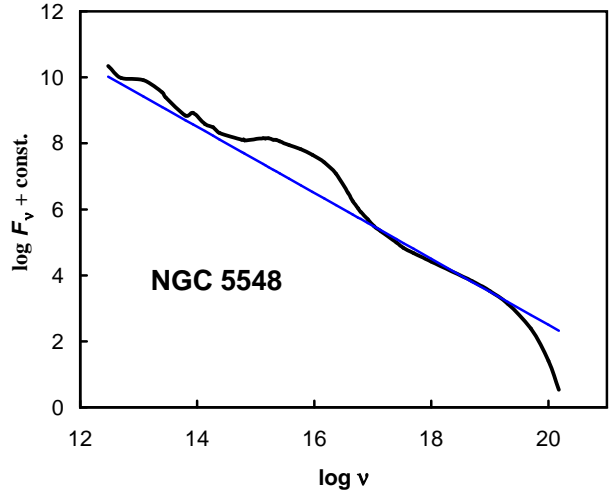}
  \caption{The spectral energy distribution of NGC 5548 (thick curve) and a
  $\nu^{-1}$ power law (thin line).  Compare this representation of the data with Fig.\@ 2 below.  (Data from
  Gaskell, Klimek, and Nazarova 2007).}
  \label{fig:fig1}
\end{figure}

\begin{figure}
  \includegraphics[width=\columnwidth]{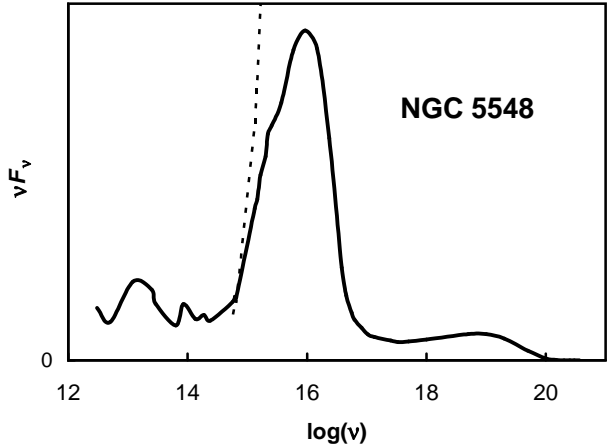}
  \caption{The spectral energy distribution of NGC 5548 (solid line)
  and the spectrum of a $F_{\nu} \propto \nu^{+1/3}$ accretion disk (dotted
  line).  Compare this representation of the observed SED with Fig.\@ 1 above.
  (SED from Gaskell, Klimek, and Nazarova 2007).}
  \label{fig:fig2}
\end{figure}

\section{Theory of AGN Energy Generation}

Once the nature of the compact sources of energy in AGNs became
apparent it was quickly recognized that the most likely source of
energy was accretion onto supermassive black holes (Zel'dovich \&
Novikov 1964, Salpeter 1964).  The theory of this is straight
forward.  As matter falls in from infinity the gravitational
potential energy ($PE$) is converted into bulk kinetic energy ($KE$)
of motion. For a bound system in equilibrium the virial theorem
tells us that $KE = -PE /2$.  So one half of the potential energy
must be lost.  This energy is lost mechanically (as mass outflow)
and via radiation.  Since the infalling matter will have non-zero
angular momentum, the virialization will take place in a disc
(Lynden-Bell 1969).  This disk was at first thought to be
geometrically thin (Pringle \& Rees 1971, Shakura \& Sunyaev 1972),
and such disks have been widely studied, but more detailed
consideration of the processes going on in disks (see Blandford \&
Begelman 2004) shows that convection and mass loss cause a
substantial thickening of the disk.  This is seen in detailed
simulations of accretion (see, for example, Stone, Pringle, \&
Begelman 1999 and Hawley \& Krolik 2001).  The resulting outflows
are well established observationally (see Crenshaw, Kraemer, \&
George 2002) and there is abundant evidence that the accretion
structures in AGNs have a substantial covering factor (see Gaskell,
Klimek, \& Nazarova 2007), and are hence geometrically thick.

To find the overall SED of an accretion disk one integrates the
emission over radius.  Since in AGNs we expect the disk to be
optically thick everywhere,\footnote{AGNs differ from X-ray binaries
in this regard because the disks in AGNs are many orders of
magnitude larger.} for our present purposes we can approximate the
local spectrum at each radius as a black body depending only on the
temperature at that radius.  With this approximation, the shape of
the overall SED then depends only on the radial dependence of the
temperature.  We will assume that the luminosity, $L$, at a given
radius, $R$, is proportional to the energy production rate at that
radius. This energy production rate per square centimeter in a ring
of thickness $dR$ at radius $R$ is proportional to the PE released
per gram of material going from $R + dr$ to $R$ and the mass
accretion rate, $dM/dt$, divided by the area of the ring of
thickness $dR$. If we put all this together we get
\begin{equation}
  L \propto \frac{GM}{R^2} dR \frac{dM}{dt}\frac{1}{\pi r dR} \propto
  \frac{1}{R^3}.
\end{equation}

Since we are assuming a black-body spectrum, $L \propto T^4$,
\begin{equation}
  T \propto L^{1/4} \propto (R^3)^{1/4} \propto R^{-3/4}.
\end{equation}

\noindent and because the wavelength, $\lambda_{max}$, of the peak
of a black body spectrum, is proportional to $T^{-1}$, most of the
radiation at a given wavelength, $\lambda$, comes from near a
radius:
\begin{equation}
  R \propto \lambda^{4/3}.
\end{equation}

To get the integrated spectrum from the disk one adds up all the
Planck curves from each radius.  It can be shown that if $T \propto
R^p$ then
\begin{equation}
  F_{\nu} \propto \nu^{(3-2/p)}.
\end{equation}

\noindent (See Pringle \& Rees 1971 or Shakura \& Sunyaev 1972).  If
$p = 3/4$ we get
\begin{equation}
  F_{\nu} \propto \nu^{+1/3}.
\end{equation}

\noindent Note that this result arises in a straight forward way
solely from considerations of energetics and mass conservation; one
does not need to worry about the much more vexing issues of how
angular momentum is transported outwards.  It also does not depend
on the thickness of the disk. The high-energy end of the spectrum is
a Wien exponential cutoff corresponding to the temperature of the
inner radius of the disk, and the low-energy limit of the spectrum
is a Rayleigh-Jeans tail of the emission from outermost part of the
disk.

Eq.\@ (5) is a well-known result, and there have been many attempts
to model the BBB in AGN SEDs with a $\nu^{+1/3}$ disk spectrum and
with additional refinements such as more realistic local emergent
spectra, the addition of Componizing coronae, and consideration of
inclination effects (see Bonning et al.\@ 2007 for a recent
example). These additional refinements mostly affect only the high
energy shape of the SED, and extend the BBB to somewhat higher
energies (see, for example, Czerny \& Elvis 1987, Laor \& Netzer
1989, or Hubeny et al.\@ 2001).

Despite the considerable effort expended on them, I believe that
{\it one should not be obsessed with the $\nu^{+1/3}$ spectra!}
First, despite impressions one might get from the literature, a
$\nu^{+1/3}$ spectra is {\it not} a good fit to an AGN SED.  We
simply {\it never} observe rising $\nu^{+1/3}$ spectra.  The only
way to get a $\nu^{+1/3}$ spectrum to fit a real AGN SED is to
arbitrarily add in a substantial power-law contribution (see, for
example, Fig.\@ 2 of Malkan 1983) or to fit a limited region of the
observed SED with the high-energy turnover in the theoretical
spectrum.\footnote{Note that the inner disk temperatures in the fits
of Malkan 1983 are more than an order of magnitude too low to
explain the extreme-UV to soft X-ray SED.} In doing this,
fine-tuning of parameters is needed and this approach gives a
fundamental problem: how does an AGN manage to convert most of its
energy into a mysterious non-thermal power law? The difficulty of
fitting $\nu^{+1/3}$ to any part of an SED is illustrated in Fig.\@
2 where the steep dotted line is $\nu^{+1/3}$.

The second reason one should not get obsessed with a $\nu^{+1/3}$
spectrum is that it is easy to get a different slope from an
accretion disk because the slope of the emergent spectrum depends
strongly on the radial temperature index, $p$. Even a small
deviation of $p$ from 0.75 will have a big effect on the slope of
the spectrum.  The important assumption in our derivation of Eq.\@
(4) is that the energy at a given radius is generated locally at
that radius by the accretion. In reality this will not be the case.
The inner part of the disk is much hotter than the outer part of the
disk (Eq.\@ 2), heat flows from hot to cold, and the effect of any
radial transfer of heat will therefore be to soften the radial
temperature gradient. The limiting case of this is to make the
opposite assumption from above, and to assume that the heating at a
given radius is not the result of local energy generation from
accretion, but comes instead from heat flow from hotter material at
smaller radii.  If we assume a single central energy source then the
flux falls off as $R^{-2}$ and the effective equilibrium temperature
falls off as
\begin{equation}
  T  \propto R^{-1/2}
\end{equation}

\noindent This give $p = -0.5$ and from Eq.\@ (3) we get a radically
different spectrum of $F_{\nu} \propto \nu^{-1}$.  An example of a
model giving a $\nu^{-1}$ spectrum is the so-called ``slim disk'' of
Abramowicz et al.\@ (1988).

Since the theory of accretion disks is still at a rudimentary stage,
I believe that it is better to turn the problem around and derive
the radial temperature distribution empirically from the observed
spectral slope rather than trying to force a theoretical spectrum
coming from a preconceived idea of the temperature structure to fit
the observations. Fig.\@ 3 shows the distribution of UV to optical
spectral indices, $\alpha_{UVO}$, for a sample of AGNs.  Note, after
allowing for the opposite sign convention for $\alpha$ in the
Gaskell et al. (2004), that nowhere in this figure do we see
anything remotely approaching $F_{\nu} \propto \nu^{+0.33}$!
Instead, it can be seen that for high luminosities there is a sharp
peak in the distribution at $\alpha_{UVO} = -0.5$. The narrowness of
this distribution is all the more striking since the UV and optical
observations (see Malkan 1984) were non-simultaneous. As one goes to
lower luminosities it can be seen that $\alpha_{UVO} = -0.5$
represents the limit of the bluest spectra. Gaskell et al.\@ (2004)
interpret $\alpha_{UVO} = -0.5$ as the intrinsic unreddened slope of
the spectrum, and interpret steeper values of $\alpha_{UVO}$ as the
result of increasing reddening as one goes to less luminous objects.
Support from this picture comes from emission-line ratios.  If we
accept the interpretation that $\alpha_{UVO} = -0.5$ is the true
slope of the spectrum in the optical and UV, then from Eq. (3) this
implies that $p = 0.57$.

$p = 0.57$ falls comfortably between our limiting cases of $p =
0.75$ for local energy generation (e.g., the standard thin disk) and
$p = 0.5$ for non-local central energy generation.  $p= 0.57$
explains the rise of the spectrum in Fig.\@ 2 as one goes from the
optical to the UV.  Looking at the areas under the solid curve in
Fig.\@ 2 shows that there we can put strong limits on any power-law
contribution.  The main energy unaccounted for is the energy in the
IR, and this is easy to explain by thermal reprocessing in the
torus.  In the thermal dust emission model the ratio of areas under
the IR portion of the curve to the area under the curve from the
optical region to higher frequencies gives the covering factor of
the dusty torus.  IR variability strongly supports the correctness
of this dust-reprocessing picture.

\begin{figure}[!t]
  \includegraphics[width=\columnwidth]{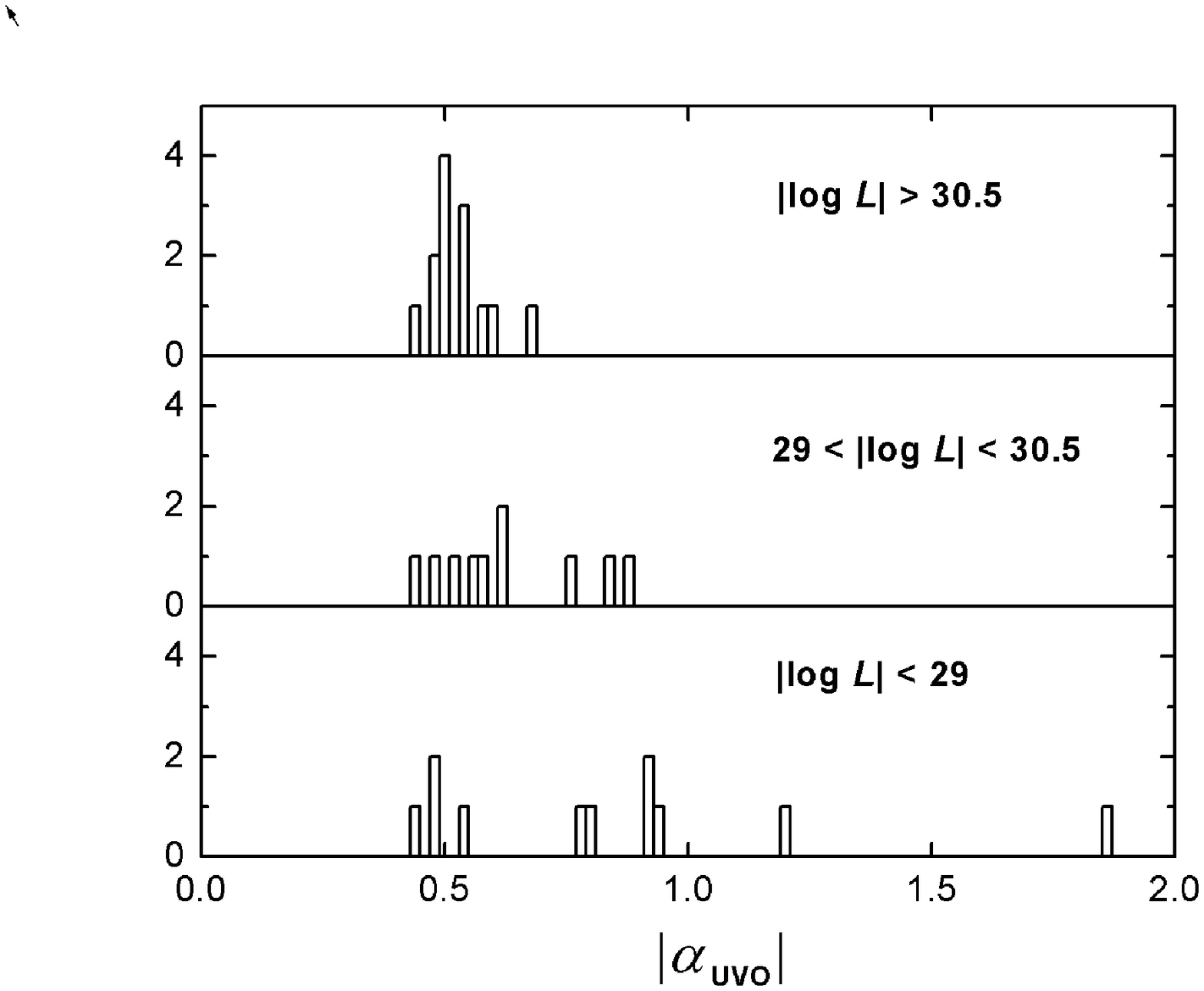}
  \caption{The distribution of optical to UV spectral indices in AGNs.
  Note that this figure, taken from Gaskell et al.\@ (2004), uses
  the opposite sign convention for $\alpha$ from this paper.}
  \label{fig:fig3}
\end{figure}

\section{Variability Fundamentals}

Variability is the change in some quantity with time.  The first two
basic questions one ought to ask about anything varying are, ``how
much does it change?'' and ``how rapidly does it change?''  If one
looks at the extensive astronomical literature on time variability
of anything one finds that the overwhelming focus is on the temporal
aspect.  One will find innumerable papers determining periods of
things or searching for periods, and there are discussions of
exponential decay times, duty cycles, power spectra, chaotic
behavior, and so on. What gets far less attention is the first
question of {\it how much} something changes,\footnote{As an
illustration of this, as of this writing, a search on the ADS for
the word ``time'' in abstracts brings up almost 200,000 hits, and
``timescale'' brings up 125,000 hits. The combination of the words
``amplitude'' and ``variability'' only generates 8000 hits.} yet
this is really the more fundamental question. It is the {\it
amplitude} which tells us how {\it important} variability really is.
A couple of examples should make this very clear.  The bolometric
amplitudes of pulsating variable stars are very small. This tells us
that this variability is unimportant in the sense that it is not the
main energy mechanism of the star which is varying. The variability
of pulsating stars is in fact a minor consequence of changing
opacity in the outer envelope of the star. Likewise with the sun,
the solar cycle has a very low amplitude.  It too has nothing to do
with the main energy generation in the core of the sun, but is a
consequence of small secondary effects to do with magnetic fields in
the outer regions of the sun.  At the other extreme, the variability
amplitude of supernovae is enormous.  This is no minor secondary
phenomena; the mechanism causing the variability in a supernova is
the mechanism causing the supernova. What about AGN variability?  Is
it a minor phenomenon like pulsations of variable stars, or is it
the fundamental energy generation mechanism of the AGN turning off
and on?

If one routinely monitors AGNs in the optical one finds that, for a
typical AGN light curve, the amplitude during the course of a year
will probably be only a couple of tenths of a magnitude or so. Since
this is less than the amplitude of a typical Cepheid variable star,
one might conclude, therefore, that, as is the case with pulsating
variable stars, variability of AGNs is just some minor phenomenon
unrelated to the main energy production I have outlined in the
previous section. However, the way observations are almost always
presented is misleading, because AGN light curves are almost
invariably shown including the star light from the nucleus of the
host galaxy. Fig.\@ 4 shows the effect of removing host galaxy
light. If one ignores the host-galaxy contamination, the variability
shown in Fig.\@ 4 is $\sim \pm 15$\%. However, after appropriate
subtraction of the host-galaxy light, it can be seen that the
amplitude of variability is almost an order of magnitude over a
six-week period.

\begin{figure}[!t]
  \includegraphics[width=\columnwidth]{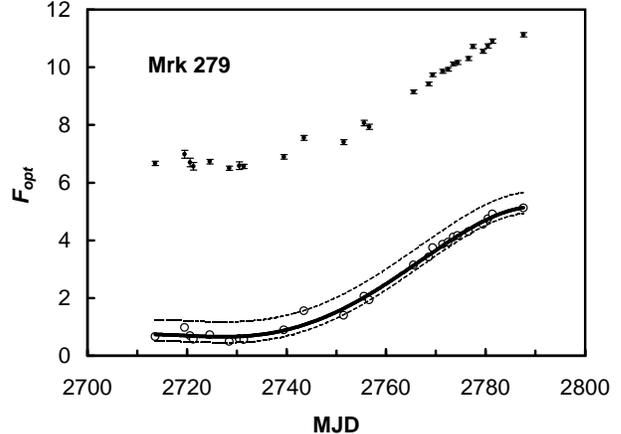}
  \caption{The V-band light curve of Mrk 279.  Top points (filled circles with
  error bars) are the observed relative fluxes observed through an 8-arcsecond photometric aperture.
  Bottom points (open circles) show the fluxes after correction for host
  galaxy light.  The solid two lighter lines on either side of the solid curve
  show the effects of the upper limits in the uncertainty in the host galaxy light correction. From
  Gaskell et al.\@ (in preparation); see also Costantini et al.\@ (2006).}
  \label{fig:fig4}
\end{figure}

\section{Sizes and Timescales}

We have seen in \S 3 that a modest modification of the radial
temperature distribution of a accretion flow enables us to explain
the single-epoch (or steady-state) main power-producing part of the
AGN spectrum.  What predictions can we make for the {\it
variability} of this main component of the spectrum?  Physically, a
timescale comes from a length scale and a relevant velocity.  I have
attempted in Table 1 to give a reasonably observer-friendly
compilation of relevant sizes, velocities, timescales, and
temperatures for a disk with the characteristics of NGC~5548 where
we believe that there is a $6.7 \times 10^8 M_{\sun}$ black hole
accreting at $L/L_{Edd} \sim 0.06$ (Koratkar \& Gaskell 1991,b;
Peterson et al.\@ 2004). Two sets of temperatures are shown.  The
first, $T_{0.72}$, has been scaled from the temperatures of the
Hubeny et al.\@ (2001) disk models (where $p = 0.72$) using the
standard $T \propto M^{-1/4} (L/L_{Edd})^{1/4}$ scaling. The second
set of temperatures, $T_{0.57}$, uses the $T \propto r^{-0.57}$
dependency deduced above from the optical--UV spectrum, and has been
scaled to produce approximately the same $L_{bol}$ as the Hubeny et
al.\@ (2001) disk. The derivations of the remaining quantities
should be obvious.

\begin{table*}
\centering
  \tablecols{11}
  \caption{Disk characteristics for NGC 5548}
  \begin{tabular}{rrcrrrrrrr}
\toprule \multicolumn{1}{c}{~~R/$R_g$}  &
\multicolumn{1}{c}{$v_{orb}$}    &   Region     &
\multicolumn{1}{c}{$R$}   &  \multicolumn{2}{c}{$t_{orb}$}         &
$T_{0.72}$    &
$T_{0.57}$   &   \multicolumn{1}{c}{$\lambda_{0.72}$}   & \multicolumn{1}{c}{$\lambda_{0.57}$}     \\
\cmidrule{5-6}
&   (km s$^{-1})$    &       &   (lt-d)    &   (days)    &   (yrs) &   (K) &   (K)      & \multicolumn{1}{c}{(\AA\@ etc.)}  &   \multicolumn{1}{c}{(\AA\@ etc.)} \\
\midrule
3   &   120,000  &       &   0.02    &   0.4    &       &   155000  &   149000  &   200 &   210 \\
10  &   70,000   & K$\alpha$      &   0.08    &   2    &       &   65000   &   75000   &   500 &   430 \\
30  &   40,000   & K$\alpha$      &   0.25    &   11   &       &   29000   &   40000   &   1100    &   800 \\
100 &   20,000   &       &   0.8    &   70   &       &   12300   &   20000   &   2600    &   1600    \\
300 &   12,000   &   BLR    &   2    &   360  &   1    &   5600    &   10800   &   6000    &   3000    \\
1000    &   7,000    &  BLR     &   8   &       &  6    &   2400    &   5400    &   1.3\,$\mu$m   &   5900    \\
3000    &   4,000    &  BLR     &   23   &       &   30    &   1100    &   2900    &   2.9\,$\mu$m  &   1.1\,$\mu$m  \\
10000   &   2,000    &  dust     &   80  &       &   200    &   400 &   1500    &   8\,$\mu$m   &   2.1\,$\mu$m   \\

    \bottomrule
  \end{tabular}
\end{table*}

The foremost thing to notice from Table 1 is how {\it big} the disk
is!  The inner and outer disk temperatures and radii are in good
agreement with the cutoff of the BBB and the dust radius
respectively for NGC~5548 (see Gaskell, Klimek, \& Nazarova 2007),
but notice how {\it the disk regions contributing to the
photoionization of the BLR come right up to and into the inner BLR}!
Notice also that the region dominating the visible light is at the
same radius as the inner BLR. The idea of the BLR overlapping with
the energy-producing region has already long been considered by Suzy
Collin-Souffrin and collaborators (see for example Collin-Souffrin,
Hameury, \& Joly 1988). The approximate correctness of the radii in
Table~1 is supported by the disk temperature falling below the dust
sublimation temperature just where the hottest dust is indeed
observed in NGC~5548 (see Gaskell et al.\@ 2007).

Table 1 also gives sizes in light-travel units, and orbital
timescales for the proposed NGC~5548 disk. It is important to note
that the {\it each spectral region is predominantly emitted at the
radius given}. This radius comes from Eq.\@ 3. Taking the optical as
an example, while there is some emission from the Rayleigh-Jeans
tail of the inner disk, the bulk of the emission comes from material
at a temperature of $\sim 6000$ K.

Czerny (2006) has given a good summary of the timescales of standard
thin accretion disks. In Table 1 I have given just the two shortest
timescales. The corollary of the large radii in Table 1 is that even
these timescales are long.  Additional timescales, such as the
sound-crossing times (see Czerny 2006) are so long for regions
producing the optical and UV radiation, that they are clearly
irrelevant for observed AGN variability and do not need to be
considered further. The dynamical (orbital) timescale, however, is
important since this is the fastest one can move large amounts of
matter into an AGN at a given radius. From Table 1 it can be seen
that it is quite long (year to decades) for the regions producing
the optical continuum. AGNs vary considerably faster than this (on a
timescale of days for NGC~5548), so we can say that, contrary to
popular opinion, {\it normal optical variability is not the result
of variations in the accretion rate on dynamical timescales.}

The only remaining relevant timescale is the light-crossing time,
and this is indeed the shortest timescale on which AGNs show strong
variability. This simple but important fact tells us that {\it
whatever causes AGN variability propagates at the speed of light, or
close to it.}

\section{The Wavelength-Dependence of AGN Variability}

At least in the steady-state approximation we have been considering
so far, different spectral regions of the SED are emitted at
different radii. Connections between wavebands therefore tell us how
energy is propagating radially, and an enormous amount of
observational effort has justifiably been put, and continues to be
put, into multi-wavelength observing. The two most common goals of
these multi-wavelength observing campaigns are (i) to search for
correlations between the variability in different wavebands, and
(ii) to search for time lags between bands. Correlations indicate
possible causal relationships, and since effects follow causes, time
lags can help us understand what causes what.

One of the most commonly investigated hypotheses is {\it
reprocessing} of radiation.  The three main ways reprocessing can
happen are (i) heating and thermal re-radiation, (ii) reprocessing
through atomic processes (emission lines and bound-free continua),
and (iii) Compton scattering.  Of these the first two produce
lower-energy photons, while Compton upscattering will produce higher
energy photons. The best understood AGN variabilty is IR
variability.  There is excellent evidence that this is the result of
reprocessing of higher energy radiation.   It has long been observed
that the IR emission follows variability at shorter wavelengths
(Clavel, Wamsteker, \& Glass 1989, Oknyanskii et al.\@ 1999, Glass
2004, Suganuma et al.\@ 2006) and that the lags are consistent with
thermal reprocessing in the dusty torus (Barvainis 1992, Gaskell,
Klimek, \& Nazarova 2007).

It is believed that the hard X-ray emission ($> 1-2$ keV is produced
by Compton upscattering of lower-energy photons. Recent overviews of
X-ray variability and its connection with variability at other
wavelengths can be found elsewhere (e.g., Gaskell \& Klimek 2003,
McHardy et al.\@ 2006, Gaskell 2006, and Uttley 2006).

It is often suggested in both the literature and in observing
proposals that {\it optical} variability is a consequence of
re-processing of X-ray emission, but it can be seen from Fig.\@ 2
that this cannot be because there is more energy in the optical/UV
than in the X-ray region. The small lags that are sometimes seen
between hard X-ray emission and the optical are probably the result
of both lagging behind the EUV/soft X-ray variability.

\section{Reprocessing in the Big Blue Bumps}

It has long been known that variability of what we now recognize as
BBB emission gets smoother and of lower amplitude as one goes to
lower energies. This is illustrated in Fig.\@ 5.  The smoothing due
to reprocessing of higher-energy radiation is a natural explanation
of this, but a long-standing problem for processing within the BBB
(as opposed to IR and X-ray reprocessing) has been that the observed
delays between the optical and the UV have been very short and were
not detected in the first {\it International AGN Watch} monitoring
campaigns (Clavel et al.\@ 1991, Peterson et al 1991, Edelson et
al.\@ 1996). In fact, we only have one clear cut case of an
optical--UV delay, NGC~7469 (Wanders et al.\@ 1997; Collier et al.\@
1998).  In Fig.\@ 6 I have rescaled and replotted the optical and UV
light curves from these two papers to show the lag.  A little
further consideration shows that reprocessing appears to work very
well in this case.  As can be seen in Fig.\@ 6, the optical light
curve is fit very well with the average of the previous five days
(to give a total delay of 2.5 days). This boxcar smoothing of the UV
light curve also reproduces the UV-optical cross-correlation
function (see Gaskell \& Sparke 1986) well from the UV
autocorrelation function.  I have shown normalized variability in
Fig.\@ 6 because there are uncertainties over the corrections for
host galaxy light and reddening, but the amplitude of the optical
variations in energy units probably does not exceed the amplitude of
the UV variability.

\begin{figure}[!t]
  \includegraphics[width=\columnwidth]{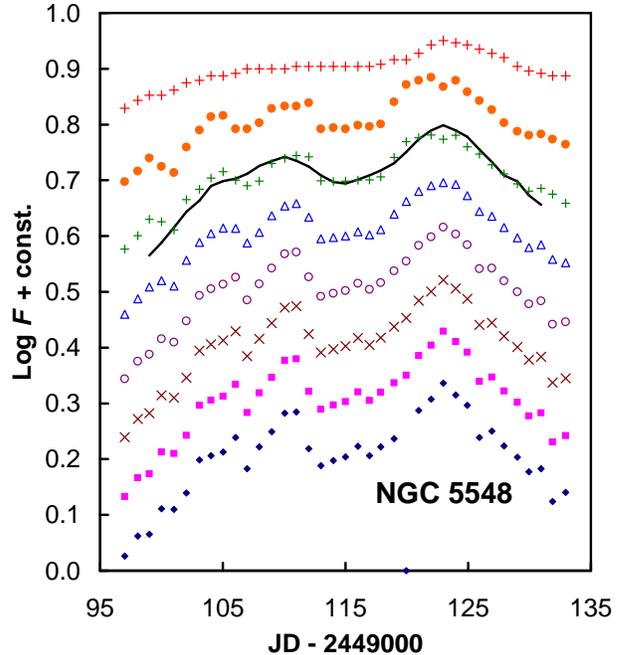}
  \caption{Light curves for NGC~5548.  Data from Korista et al.\@ (1995).
  The light curves, starting from the bottom are, at
  $\lambda$$\lambda$ 912 (extrapolated), 1145, 1350, 1460, 1790, 2030, 2195, and
  5100.  Notice the decreasing amplitudes and greater smoothness of the light curves as
  one goes to to longer wavelengths.
  The smooth curve through the $\lambda$2030 curve is the $\lambda$912
  light curve smoothed with a boxcar function running from $-3$ to $+3$ days.}
  \label{fig:fig5}
\end{figure}

\begin{figure}[!t]
  \includegraphics[width=\columnwidth]{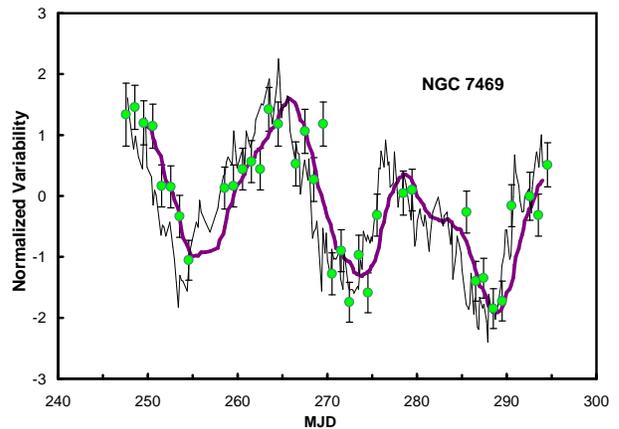}
  \caption{The relative UV, $\lambda$1315, variability of NGC~7469 (thin line) compared
  with the optical, $\lambda$4845, variability (circles with error bars).
  The thick solid curve is the UV variability smoothed with a boxcar
  function extending from 0 to 5 days.  The variations are shown on
  a linear scale normalized to the same variance.  (Data from Wanders et al.\@ 1997, Collier et
  al.\@ 1998, and Kriss et al.\@ 2000)}
  \label{fig:fig5}
\end{figure}

For NGC~7469 we found that the wavelength-dependent lags,
$\tau(\lambda)$, in NGC~7469 could be fit well by
\begin{equation}
\tau(\lambda) \propto \lambda^{4/3}
\end{equation}

\noindent (see Wanders et al.\@ 1997, Collier et al.\@ 1998, and
Kriss et al.\@ 2000).  This is consistent with Eq.\@ 3 if
$\tau(\lambda) \propto R$.

Although reprocessing looks convincing in this case, the shortness
of the delay, which is on the order of the light crossing-time, is
problematic.  We therefore interpreted the lags as a consequence of
light-travel time in the external illumination of a disk with a $T
\propto R^{-3/4}$ radial temperature structure (see Wanders et al.\@
1997, Collier et al.\@ 1998, and Kriss et al.\@ 2000). From
broad-band optical photometry going out to the near IR (0.9 $\mu$m),
Sergeev et al.\@ (2005) have found many more wavelength-dependent
lags and also shown that the lags at a given wavelength are
proportional to $L^{1/2}$ where $L$ is the optical luminosity of the
AGN. To explain this they also suggest an external illumination
source and postulate that the height of this external illumination
source depends on the square-root of the luminosity.

As discussed in Gaskell (2007) there are many problems with the
external illumination model.  Foremost among these are that the
external illumination source is never seen from the earth, and
because the amplitude of the variability can be large (see, for
example, Fig.\@ 4), this mysterious energy source, rather than the
disk itself, must be the main energy source.  Instead I have
proposed (see Gaskell 2007) that the general increase in lag Sergeev
et al.\@ find towards the near IR comes naturally as a result of
increasing contamination from re-emitted light from the very hot
dust in the torus.

The nice reprocessing picture I have shown for NGC~7469 in Figs.\@ 6
and 7 is shattered when we go back to looking at the variability of
NGC~5548 Fig.\@ 5.  he variability does indeed get smoother as one
goes to longer wavelengths and the amplitude decreases as expected
in the reprocessing model.  Again there is good quantitative
agreement illustrated both by the fit of a smoothed short wavelength
curve to one of the longer UV wavelengths in Fig.\@ 5 and to the
optical continuum autocorrelation function. But there is a {\it huge
problem with causality} -- the $\pm 3$ day boxcar function used in
Fig.\@ 5 and the wider ($\pm 4.5$ day) one needed to explain the
optical continuum autocorrelation function extend to both positive
and negative times! In other words, we need to reprocess photons
that will not be emitted until several days in the future!

The simple reprocessing picture continues to collapse when we
consider an AGN for which we have good long-term soft X-ray, UV, and
optical monitoring.  In Fig.\@ 9 I shown the soft X-ray, UV, and
optical light curves of 3C~390.3. Since 3C 390.3 has a substantial
(and as yet undetermined) host galaxy contribution. I have
subtracted an arbitrary constant flux to emphasize the optical
variations.

\begin{figure}[!t]
  \includegraphics[width=\columnwidth]{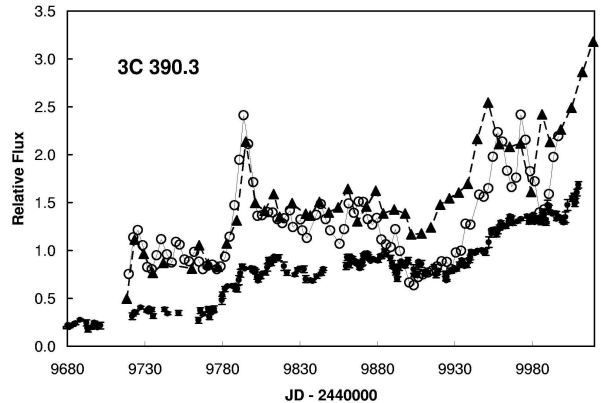}
  \caption{Soft X-ray, UV, and optical fluxes of 3C~390.3.  The open
circles are 0.1 -- 2 keV {\it ROSAT} HRI fluxes; the black triangles
are {\it IUE} $\lambda$1370 UV fluxes, and the small black circles
with error bars are V-band fluxes. The {\it ROSAT} fluxes are from
Leighly et al.\@ 1997, the UV fluxes are based on O'Brien et al.\@
(1998), and V-band fluxes have been derived with scalings and
corrections from Dietrich et al.\@ (1998). Scale factors have been
chosen for plotting convenience, and an arbitrary constant has been
subtracted from the V-band fluxes. Figure reproduced from Gaskell
(2006)}
  \label{fig:fig7}
\end{figure}

Around MJDs 9725 and 9790 we see what we would expect from
reprocessing.  A strong soft X-ray ``flare'' with an almost
simultaneous UV flare, and a much less obvious optical event.  But
that is where the agreement with reprocessing expectations ends.
Right after the MJD 9790 event there is a follow-on event in the
optical {\it of comparable magnitude to the optical event at MJD
9790}, but there are at best only very weak corresponding UV and
soft X-ray events.  A soft X-ray ``anti-flare'' (dip) at MJD 9855
has no corresponding event in the UV or optical.  Then the X-rays go
into a ``low state'' around MJD 9900 and the optical follows suit
about 10 days later (this is more obvious in the smoothed curves --
see Gaskell 2006).  But then look at what happens -- there is a {\it
major} flare in the UV. This has no counterpart in the optical, but
a major flare in the soft X-rays {\it follows} the UV flare! Another
soft X-ray major flare, fast on the heels of the first, has no
obvious counterparts in the UV and the optical.

\section{A Model for Multi-Wavelength Variability of AGNs}

Clearly the behavior of 3C~390.3 and NGC~5548 rules out the widely
considered simple reprocessing picture.  Instead it points to {\it
different spectral regions varying relatively independently}, which
is just what we expect from the temperature structure of the
accretion flow. Variability at one radius produces UV variability;
variability at another produces optical variability.  This is
energetically quite feasible.  If the effective temperature of just
a quarter of the surface area of what I will loosely call the
``optical'' region of the disk doubles, this will increase the
energy output of the region varying 16 fold and the quadruple the
energy output of the whole optical region of the disk. The
variability at this radius will propagate inwards and outwards to
hotter and cooler regions.  Given perfect multi-wavelength,
monitoring the radius varying can be identified from the part of the
SED which varies most strongly.

If the AGN is not face on but has a significant inclination angle
the lags between events at different frequencies will also depend on
whether the driving event is into or out of the plane of the sky.
The lag will be less when the disturbance in propagating towards the
observer.

This model makes quite a number of predictions that are testable at
least in practice.

(1) The closer regions of the SED are to each other in energy, the
more correlated their variability will be.  This does indeed seem to
be the case.

(2) When a disturbance originates in a cooler region giving a
low-energy event and this disturbance propagates inwards, the higher
energy event will {\it follow} the lower energy event.  This should
be testable.  As one moves away from the spectral region of maximum
variability (identified by the maximum variability) both lower
energy {\it and the higher energy} variability will lag behind the
driving variability.

(3) The timescale of driving events of comparable relative amplitude
in different spectral regions will vary as $\lambda^{\beta}$ where
$\beta = 1/p$.

(4) Secondary events will be smoothed relative to driving events,
although this might not always be detectable.

(5) Because individual variations in the ionizing continuum are no
longer axially symmetric, light echoes in emission lines and
polarized flux (see Gaskell et al. 2007) will be different for
different events, even though there is insufficient time for the
mean distance of the emission-line regions or polarizing electrons
to change their distance from the central black hole.  Although
somewhat bad news for reverberation mappers, this is quite a
powerful test, and hitherto puzzling changes in lags for separate
consecutive events have already been noticed (Maoz 1994).

(6) Changes in line lags and color-dependent lags will be greater
for AGNs with higher inclination angles because the azimuthal
location matters more.

\section{Where Next?}

I believe that there are a couple of clear directions we need to be
going in.  First, we need more data.  A lot of effort has gone into
AGN monitoring, but it has been concentrated on only a few objects.
Astrophysicists, along with vertebrate paleontologists are probably
the world leaders in building theories on the observations of a
single object! I hope it is clear from my discussion of the
contrasting multi-wavelength behavior of NGC~5548, NGC~7469, and 3C
390.3 that one can get {\it very} misleading impressions from just a
single short-duration observing campaign looking at only one object.
Yet this is where a lot of our ideas of AGN variability have come
from. We need comparable or better monitoring of quite a few
objects, and preferably ultimately of well chosen samples.

The second direction I believe we need to be going in is in making
more detailed 3D-magnetohydrodynamic simulations of accretion onto
compact objects.  I find the results obtained so far by Hawley \&
Krolik (2001) and others to be very encouraging, and this sort of
thing is only going to get easier as increases in computing power
shows no signs of tapering off.  Although treatment of all relevant
radiative processes is not yet included in these simulations, Hawley
\& Krolik (2001) find that fluctuations in the volume-integrated
Maxwell stress and accretion rate at the inner edge of the disk
(both of which are probably related to the luminosity) fluctuate
wildly (i.e., the energy generation is indeed fundamentally
unstable) and that the Fourier power spectrum of the variability,
$P(f)$, has a ``red'' $P(f) \propto f^{-1.8}$ shape similar to what
is observed in AGNs. NGC ~5548, for example, has indices of -3 and
-2.5 at different epochs (Koratkar \& Gaskell 1991, Krolik et al.\@
1991), and the Reichert et al.\@ (1994) observations of NGC~3783
give an index of -1.8 (White \& Peterson 1994).  It remains to be
seen, however, how well such simulations will be able to reproduce
the rapidity of flux variations and the small lags.

Finally, although they differ from AGNs in many ways, it is possible
that observations of accreting binaries could offer useful insights
into what is going on with AGNs.  For example, eclipse ``flicker
mapping'' of the dwarf nova V2051 Oph (Baptista \& Bortoletto 2004)
shows that there is high-frequency flickering all over the disk with
a constant relative amplitude independent of disk radius and
accretion rate.  This is consistent with the model of AGN
variability I have suggested, although the amplitude of the dwarf
nova flickering is lower (3\% rms) for V2051 Oph.

\section{Conclusions}

Variability rules out the standard popular geometrically-thin
``$\alpha$-disk'' paradigm of Pringle \& Rees (1971) and Shakura \&
Sunyaev (1972). Variability in the UV and optical is simply too fast
and much too simultaneous to be due to fluctuations in the accretion
rate in a standard thin disk.  The rapidity and simultaneity instead
require that the variability mechanism propagates close to the speed
of light.  Since the amplitude of AGN variability is enormous the
variability mechanism is the main energy generating mechanism of
AGNs.  Both theory and observations strongly point to variability
being {\it localized} in the disk, and there are tests that can be
made of this hypothesis.

\acknowledgments  I would like to thank all the organizers for
making our time in Huatulco flow so smoothly, and my many
collaborators, colleagues, and students for all their stimulation of
my thoughts on the issues discussed here. US taxpayers unwittingly
made this research and my trip to the tropical paradise of Huatulco
possible through National Science Foundation grant AST 03-07912 and
Space Telescope Science Institute grant AR-09926.01.

%
%

\end{document}